\documentclass[preprintnumbers]{revtex4}
\usepackage{amssymb}
\usepackage{amsmath}
\usepackage{graphicx}
\usepackage{calligra}
\usepackage{mathrsfs}
\usepackage{dcolumn}
\usepackage{bm}
\usepackage{subfigure}
\usepackage{color}
\usepackage{CJKutf8}
\numberwithin{equation}{section}
\numberwithin{figure}{section}

\begin{document}
\title{Photon Spheres and shadow of Schwarzschild black hole on the EUP framework}
\author{Hai-Long Zhen$^{1,2}$, Jian-Hua Shi$^{1,2}$, Huai-Fan Li$^{2,3}$, Yu-Bo Ma$^{1,2}$}
\thanks{\emph{e-mail:} \texttt{03100064@sxdtdx.edu.cn} (corresponding author)}
\affiliation{$^1$Department of Physics, Shanxi Datong University, Datong 037009, China\\
$^2$Institute of Theoretical Physics, Shanxi Datong University, Datong, 037009, China\\
$^3$College of General Education, Shanxi College of Technology, Shuozhou 036000, China\\}

\begin{abstract}

An explicit correspondence is established between the Extended Uncertainty Principle (EUP) and the metric function by directly relating the radiation temperature function modified by EUP to the modified spacetime metric. Utilising this modified metric, we subsequently derive the corresponding thermodynamic quantities of the black hole, and calculate the photon sphere radius and the size of the black hole shadow. The results of the study indicate that, in comparison with Schwarzschild black holes, the position of the event horizon remains constant under EUP modifications. However, the photon sphere radius increases with growing EUP parameters, while the shadow size decreases with increasing parameters, demonstrating that EUP induces optical shift phenomena. By comparing with observations of the galactic centre black hole $\text{Sgr}{{\text{A}}^{*}}$ from the Event Horizon Telescope, new constraints are established on EUP parameters.

\end{abstract}

\maketitle

\section{Introduction}\label{one}

Black holes represent a pivotal component in the universe. These galaxies release immense energy by accreting surrounding matter and feed back into their environment through jets, profoundly influencing the formation and evolution of galaxies. It is widely accepted that black holes represent a pivotal element in the pursuit of enlightenment concerning the fundamental enigmas of physics \cite{K9905030,R1965,C171110256}. These enigmas encompass a diverse array of subjects, including the behaviour of strong-field gravitational systems, the potential emergence of hitherto unexplored fundamental degrees of freedom, the unification of quantum mechanics and gravitational theory, and even the revelation of the nature of spacetime itself \cite{S1976,L180605195}. In the contemporary epoch, humanity has entered a new era in which it is capable of conducting routine, multi-scale observations of black holes.

On 10 April 2019, the global astronomical community simultaneously released the first image of a black hole. The black hole is located at the centre of the nearby galaxy M87, which is situated 55 million light-years from Earth. The mass of this black hole is 6.5 billion times that of the Sun. As one of the most enigmatic and destructive celestial objects in the universe \cite{K190611238,K190611239,K190611243,KL25,KL26}, the discovery of such supermassive black holes not only challenges traditional theories of black hole formation but also compels the scientific community to re-examine the fundamental nature and evolutionary mechanisms of black holes.

In the vicinity of a black hole, gravitational forces exerted by the object result in the deflection of light rays, a phenomenon known as gravitational lensing \cite{H250820419,R220602161,S220612820,K9904193,U260300916,P180100860}. This phenomenon leads to a substantial decrease in the observational intensity within the region delimited by the black hole, culminating in the formation of a low-luminosity area on the distant observation plane-the black hole shadow. The boundary of this shadow is defined by the trajectories of photons that can escape to infinity, and its geometric dimensions exceed those of the event horizon. It is noteworthy that the black hole shadow demonstrates a significant correlation with spacetime geometry; the morphology of the shadow exhibits distinct variations among diverse black hole types, thereby offering indispensable observational indicators for the investigation of black hole characteristics \cite{P241101261}. Consequently, shadow analysis has become an effective tool for estimating black hole parameters \cite{A250700455,S251207922,V947,S260401,P250904261,J251023047,W251106017,A139945,X250803020,K250208388,S230902709,M20248437,H202484350,R892782020,S201108023,M210311417,S220602488} and testing gravitational theories such as general relativity \cite{A210108261,Z211101752,V210507101,R220700078,S220507787}.

In principle, substantial advances have been made in the geometric characterisation of the photon sphere and the black hole shadow \cite{C190412983,M240803241,E240909750}. In the context of static, spherically symmetric spacetimes, the photon sphere can be delineated by the conditions of zero geodesic curvature and zero Gaussian optical curvature. This provides a coordinate-independent mathematical framework for the study of zero-geodesic circular orbits. Any deviation from the standard Bekenstein-Hawking entropy-area relation (and consequently from Schwarzschild geometry) naturally leads to modifications of the spacetime metric. Building on this principle, \cite{A251213769} systematically investigates modified metrics corresponding to different forms of entropy modification, calculates the resulting photon sphere radii and shadow dimensions, and reveals that each entropy modification induces characteristic optical shift effects. This study establishes new constraint boundaries for various entropy deformation parameters by comparing with Event Horizon Telescope (EHT) observations of the galactic centre black hole $\text{Sgr}{{\text{A}}^{*}}$.

Ref. \cite{H250820419} focuses on minimal deformation models for two types of Schwarzschild black holes: the Kazakov-Solodukhin (KS) black hole arising from quantum corrections, and the Ghosh-Kumar (GK) black hole derived from nonlinear electrodynamics (NED). It is evident that both of these approaches result in minimal deformation through straightforward algebraic substitutions of the radial coordinate in the metric function. Each approach introduces a "hair" parameter, which serves to characterize the degree of deformation. The study found that, upon analysis, the parameter in question has a significant effect on the black hole shadow. In comparison to Schwarzschild black holes, the event horizon, photon sphere radius and critical collision parameter all increased for KS black holes, while the corresponding quantities decreased for GK black holes. When considered in conjunction with EHT observational data, this work establishes constraints on the permissible parameter ranges for both types of black hole models. Meanwhile, the Lorentzian-Euclidean black hole has recently been introduced as a geodesically complete spacetime, characterized by a signature change at the event horizon that prevents causal geodesics from reaching the central singularity at $r=0$. To identify deviations of this geometry from the standard Schwarzschild solution, the shadows produced by this black hole model have been systematically investigated in Refs. \cite{E260110806,S240417267,S250708431}.

The manifestation of quantum gravity corrections in the context of black hole thermodynamics has also attracted considerable attention. Examples of such corrections include the Bekenstein-Hawking entropy correction and the extended generalized uncertainty principle (EGUP) correction to the Hawking radiation temperature. Ref. \cite{H250815467} investigates the impact of quantum corrections on the thermodynamic properties of Reissner-Nordstr\"{o}m black holes containing quintessence matter within a dark energy background. Ref. \cite{B250908005} undertakes a systematic investigation of the thermodynamic properties and $P-V$ critical behaviour of RN-AdS black holes surrounded by perfect fluid dark matter (PFDM) within the EGUP framework. Ref. \cite{Y250814953} undertakes a unified topological and geometric analysis of charged AdS black holes embedded in quintessence fields, incorporating infrared gravitational corrections introduced by the extended uncertainty principle (EUP). Ref. \cite{A250610621} investigates the accelerated radiation phenomenon exhibited by a two-level atom in free fall around a GUP-modified Schwarzschild black hole. Ref. \cite{B240515558} calculates the EGUP corrections to the thermodynamic quantities of RN-AdS black holes, including the Hawking temperature, heat capacity, entropy function and pressure, under the influence of quintessence, a dark energy candidate with negative pressure.

It is imperative to emphasise that any deviation from the Bekenstein-Hawking entropy-area relation will induce corresponding modifications to the spacetime metric. Recent studies \cite{A251104613} demonstrate that, starting from a selected entropy functional, one can systematically derive the associated modified metric and its effective matter distribution, thereby establishing a direct entropy-geometry correspondence. Inspired by this and drawing on the approach in \cite{A250721663} that derives van der Waals black holes by equating the van der Waals system temperature with the Hawking radiation temperature, this paper investigates the modified metric corresponding to the Hawking radiation temperature with EUP corrections. The impact of the EUP effect on the position of the photon sphere and the shadow radius of black holes is analysed further, based on the derived modified metric. The results indicate that as the EUP parameter increases, the photon sphere radius expands relative to that of a Schwarzschild black hole, while the shadow radius contracts; the position of the black hole's event horizon remains unchanged. This finding provides a novel perspective on the role of the uncertainty principle in shaping black hole shadow characteristics, with significant implications for deepening our cosmic understanding.

The structure of this paper is as follows: In Sec. \ref{two}, the EUP formalism is briefly reviewed. In Sec. \ref{three}, the photon sphere and shadow radius of the modified black hole based on the EUP modification metric are discussed, and the effects of various parameters on the photon sphere and shadow are analysed. Furthermore, the theoretical calculations are compared with observational data of the galactic centre black hole $\text{Sgr}{{\text{A}}^{*}}$ to provide constraints on the EUP parameters. Finally, the research conclusions are summarised and analysed in Sec. \ref{four}.

\section{Heisenberg uncertainty principle} \label{two}

By comparing observations of $\text{Sgr}{{\text{A}}^{*}}$ with the Event Horizon Telescope, constraints are provided on the parameters introduced in the Extended Uncertainty Principle (EUP). This approach provides a viable avenue for investigating the effects of generalized uncertainty relations on spacetime structure.

The expression for the EUP-modified Hawking temperature is given by \cite{H250815467,B250908005,Y250814953,A250610621,B240515558}
\begin{align}\label{2.1}
{{T}_{EUP}}&=4r_{+}^{2}\left( \frac{1}{4r_{+}^{2}}+\frac{a}{{{L}^{2}}} \right){{T}_{+}}={{\left. {{f}_{EUP}}(r,\alpha ) \right|}_{r={{r}_{+}}}}{{T}_{+}},
\end{align}
where $\alpha $ is the dimensionless EUP parameter of order one, $L$ signifies the large-scale length, and ${{T}_{+}}$ represents the Hawking temperature.

The Schwarzschild spacetime, in coordinates $(t,r,\theta ,\varphi )$,  is described as follows:
\begin{align}\label{2.2}
d{{s}^{2}}&=-fd{{t}^{2}}+{{f}^{-1}}d{{r}^{2}}+{{r}^{2}}(d{{\theta }^{2}}+{{\sin }^{2}}\theta d{{\varphi }^{2}}),
\end{align}
where $f=1-\frac{2M}{r}$, the event horizon position ${{r}_{+}}$ of the black hole is determined by the condition $f({{r}_{+}},M)=0$. The corresponding Hawking temperature is given by
\begin{align}\label{2.3}
{{T}_{+}}&=\frac{f'({{r}_{+}})}{2\pi }=\frac{1}{8\pi {{r}_{+}}}.
\end{align}

Using the Eqs. (\ref{2.1}) and  (\ref{2.3}), the black hole entropy can be derived from the first law of thermodynamics, $dM=TdS$, which yields \cite{H250815467,B250908005,Y250814953,A250610621,B240515558},
\begin{align}\label{2.4}
{{S}_{EUP}}&=\int{\frac{dM}{{{T}_{EUP}}}}.
\end{align}

Performing the integration via a Taylor expansion in the small $\alpha $ leads to
\begin{align}\label{2.5}
{{S}_{EUP}}&=\int{\frac{dM}{{{T}_{EUP}}}}={{S}_{BH}}\left( 1-\frac{2\alpha r_{+}^{2}}{{{L}^{2}}} \right)+o({{\alpha }^{2}}).
\end{align}
where the integration constant has been fixed such that the Bekenstein-Hawking entropy ${{S}_{BH}}=\pi r_{+}^{2}$ is recovered in the limit $\alpha \to 0$.

In recent years, studies have demonstrated that Anti-de Sitter (AdS) black holes exhibit behavior analogous to that described by the van der Waals equation, with their parameters undergoing first-order or second-order phase transitions within specific ranges \cite{D12050559,R13066233,R160608307,J150201428}. Notably, references \cite{A250721663} constructed spacetime line elements satisfying the condition that the radiation temperature of the black hole equals the van der Waals temperature, giving rise to what are termed van der Waals black holes. Furthermore, these works derived spacetime line elements under the scenario where particles satisfy the uncertainty relation. This line of inquiry provides a foundation for further exploration of van der Waals black holes in a variety of spacetime contexts. Drawing inspiration from this body of work, the present study investigates how the radiation temperature, modified by the Heisenberg uncertainty principle (HUP), manifests in spacetime line elements.

The EUP-corrected Schwarzschild spacetime, in coordinates $(t,r,\theta ,\varphi )$, is written as
\begin{align}\label{2.6}
d{{s}^{2}}&=-{{f}_{\alpha }}(r,\alpha )d{{t}^{2}}+f_{\alpha }^{-1}(r,\alpha )d{{r}^{2}}+{{r}^{2}}(d{{\theta }^{2}}+{{\sin }^{2}}\theta d{{\varphi }^{2}}),
\end{align}

First, the function ${{f}_{\alpha }}(r,\alpha )$ must satisfy
\begin{align}\label{2.7}
{{\left. \frac{d{{f}_{\alpha }}(r,\alpha )}{4\pi dr} \right|}_{r={{r}_{+}}}}&={{T}_{EUP}},
\end{align}
with ${{r}_{+}}$ denoting the event horizon position of black hole, and ${{T}_{EUP}}$ given by the Eq. (\ref{2.7}).

Second, the event horizon position of black hole ${{r}_{+}}$ is determined by the condition ${{f}_{\alpha }}({{r}_{+}},\alpha )=0$. Under these two constraints conditions, the following ansatz is adopted
\begin{align}\label{2.8}
{{f}_{\alpha }}(r,\alpha )&={{f}_{EUP}}(r,\alpha )f(r),
\end{align}
where
\begin{align}\label{2.9}
f(r)&=1-\frac{2M}{r},~~{{f}_{EUP}}(r,\alpha )=\left( 1+\frac{4\alpha {{r}^{2}}}{{{L}^{2}}} \right).
\end{align}

When $\alpha \ge 0$, Eq. (\ref{2.9}) implies that ${{f}_{EUP}}(r,\alpha )=0$ has no real roots. The event horizon position ${{r}_{+}}$ is determined by the condition
\begin{align}\label{2.10}
{{f}_{\alpha }}({{r}_{+}},\alpha )&={{f}_{EUP}}({{r}_{+}},\alpha )f({{r}_{+}})=0.
\end{align}
Since ${{f}_{EUP}}(r,\alpha )=0$ admits no real solutions, the condition reduces to $f({{r}_{+}})=0$. Consequently, the event horizon position ${{r}_{+}}$ and the corresponding spacetime structure remain unaffected by the EUP in the regime $\alpha \ge 0$.

Let us now use the metric Eq. (\ref{2.8}) to calculate the components of the Einstein tensor $G_{\nu }^{\mu }$. We find \cite{A250917630}
\begin{align}\label{2.11}
G_{t}^{t}&=\frac{12a}{{{L}^{2}}}-\frac{16a}{{{L}^{2}}}\frac{M}{r}=G_{r}^{r},
\end{align}
\begin{align}\label{2.12}
G_{\theta }^{\theta }&=\frac{12a}{{{L}^{2}}}-\frac{8a}{{{L}^{2}}}\frac{M}{r}.
\end{align}

When the EUP correction is taken into account, i.e., $\alpha \ne 0$, we have $G_{t}^{t}=G_{r}^{r}\ne 0$ and $G_{\theta }^{\theta }\ne 0$. Consequently, from field equations of general relativity
\begin{align}\label{2.13}
G_{\nu }^{\mu }&=8\pi T_{\nu }^{\mu },
\end{align}
we conclude that a non-zero, effective stress-energy tensor of EUP origin emerges. In particular, we may parameterize $T_{\nu }^{\mu }=(-\rho ,{{p}_{r}},{{p}_{t}},{{p}_{t}})$, with
\begin{align}\label{2.14}
\rho &=-\left( \frac{12a}{{{L}^{2}}}-\frac{16a}{{{L}^{2}}}\frac{M}{r} \right)=-\frac{12a}{{{L}^{2}}}\left( 1-\frac{4M}{3r} \right),~~\rho =-{{p}_{r}},~~{{p}_{t}}=\frac{12a}{{{L}^{2}}}-\frac{8a}{{{L}^{2}}}\frac{M}{r}
\end{align}

Notably, the relation $\rho \ne {{p}_{r}}$ reflects the fact that the equation of state in the radial direction exhibits characteristics similar to those of vacuum energy or dark energy. However, since ${{p}_{r}}\ne {{p}_{t}}$, the effective matter sector originating from the EUP manifests anisotropy. Consequently, the deviation of the radiation temperature from the standard Hawking temperature is a geometric source, generating an anisotropic stress energy tensor without introducing any explicit matter fields. In this sense, the EUP's modification of the Hawking radiation temperature of a black hole, via the first law of thermodynamics, further leads to a modification of the black hole entropy. This modification of entropy is manifested as an effective and emergent gravitational matter content associated with the microscopic structure of the event horizon. This finding is consistent with the entropy modification results obtained in Ref. \cite{A251104613}.

\section{Photon Spheres and shadow} \label{three}

A photon sphere is defined as a region consisting of null geodesics that form the boundary for any stable photon orbits within the strong gravitational field of a black hole. The concept signifies the location where light undergoes extreme bending due to the intense curvature of spacetime. In this section, the primary focus is on the impact of EUP on black hole photons.

The motion of a massless photon along its geodesics is governed by the action of the Lagrange density
\begin{align}\label{3.1}
H&=\frac{1}{2}{{g}^{\mu \nu }}{{p}_{\mu }}{{p}_{\nu }}=0,
\end{align}
where ${{p}_{\mu }}=\frac{d{{x}_{\mu }}}{d\lambda }$ denotes the generalized and momenta $\lambda $ is the affine parameter. In view of the spherically symmetric nature of the system, the focus shall now be directed towards equatorial geodesics with $\theta =\pi /2$. Consequently, the associated Lagrange function becomes
\begin{align}\label{3.2}
-{{f}_{\alpha }}(r,\alpha ){{\left( \frac{dt}{d\tau } \right)}^{2}}+f_{\alpha }^{-1}(r,\alpha ){{\left( \frac{dr}{d\tau } \right)}^{2}}+{{r}^{2}}{{\left( \frac{d\varphi }{d\tau } \right)}^{2}}&=0,
\end{align}
where $\tau $ denotes the proper time. Furthermore, two conserved quantities of the static and spherically symmetric spacetime are identified
\begin{align}\label{3.3}
E&=-\frac{\partial H}{\partial \dot{t}}=f(r,\alpha )\dot{t},~~\bar{L}=-\frac{\partial H}{\partial \dot{\varphi }}={{r}^{2}}\dot{\varphi },
\end{align}

Utilising the aforementioned equations, the following equation is derived \cite{H250820419}
\begin{align}\label{3.4}
{{V}_{eff}}&=\frac{\sqrt{f(r,\alpha )}}{r}.
\end{align}

The Lyapunov exponent of the null geodesic is as follows: \cite{S230811883,X231211912}
\begin{align}\label{3.5}
\lambda &=\pm \sqrt{\frac{V_{eff}^{''}}{2{{{\dot{t}}}^{2}}}},
\end{align}
where the prime stands for the derivative of the radial radius, denoted by $r$. The stability of the photon circular orbital can be investigated by examining the second derivative of the effective potential, under the conditions that ${{V}_{eff}}=0$ and $V_{eff}^{'}=0$. It is evident that the signs of the Lyapunov exponent are unrelated to the stability of the photon's circular orbit. Therefore, it is advisable to select a positive value for the Lyapunov exponent in the subsequent analysis.

Utilising Eqs. (\ref{3.2}) and (\ref{3.3}), the orbit equation for the massless photon is derived,
\begin{align}\label{3.6}
{{\left( \frac{dr}{d\varphi } \right)}^{2}}&={{r}^{4}}\left( \frac{{{E}^{2}}}{{{{\bar{L}}}^{2}}}-\frac{{{f}_{\alpha }}(r,\alpha )}{{{r}^{2}}} \right).
\end{align}

It is evident that the orbit equation depends on a single constant of motion, i.e., the impact parameter $b=\bar{L}\text{/}E$. It should be noted that the aforementioned equation takes the same form as an energy conservation law in one-dimensional classical mechanics, as illustrated in Eq. (\ref{3.3}), where the effective potential also depends on the impact parameter, with the coordinate $\varphi $ playing the role of the time variable. The utilisation of a fixed impact parameter facilitates the visualisation of radial motion through the employment of classical potential. In this context, the radius of the unstable photon orbit, denonted by ${{r}_{ps\alpha }}$, is determined implicitly by
\begin{align}\label{3.7}
{{r}_{ps\alpha }}f'({{r}_{ps\alpha }}){{f}_{EUP}}({{r}_{ps\alpha }},\alpha )-2f({{r}_{ps\alpha }})&=0,
\end{align}

From Eq. (\ref{3.7}), we obtain
\begin{align}\label{3.8}
{{r}_{ps\alpha }}&=\frac{1\pm \sqrt{1-48\alpha {{M}^{2}}/{{L}^{2}}}}{8\alpha }{{L}^{2}},
\end{align}

The corresponding shadow radius is given by
\begin{align}\label{3.9}
{{b}_{ps\alpha }}&=\frac{{\bar{L}}}{{{E}_{c}}}={{\left. \frac{r}{\sqrt{f(r,\alpha ,0)}} \right|}_{{{r}_{ps\alpha }}}}=\frac{{{r}_{ps\alpha }}}{\sqrt{\left( 1-\tfrac{2M}{{{r}_{ps\alpha }}} \right)\left( 1+\tfrac{4\alpha r_{ps\alpha }^{2}}{{{L}^{2}}} \right)}},
\end{align}

From Eq. (\ref{3.8}), the condition $0<48\alpha {{M}^{2}}/{{L}^{2}}\le 1$ must hold. Introducing the parameter $\alpha =\chi \frac{{{L}^{2}}}{48{{M}^{2}}}$, with $0\le \chi \le 1$, Eq. (\ref{3.8}) becomes ${{r}_{ps\alpha }}=6M\frac{1\pm \sqrt{1-\chi }}{\chi }$. In the limit $\chi =1$, we recover ${{r}_{ps\alpha }}=6M$. As $\chi \to 0$, the Eq. (\ref{3.8}) with the "$-$" yields ${{r}_{ps}}=3M$, corresponding to the standard photo sphere radius in Schwarzschild spacetime. The "$+$", however, leads to ${{r}_{ps\alpha }}\to \infty $ in the same limit, which is physically meaningless. Consequently, the "$-$" is retained and the following is written:
\begin{align}\label{3.10}
\frac{{{r}_{ps\alpha }}}{3M}&=2\frac{1-\sqrt{1-\chi }}{\chi }.
\end{align}

For a constant value of $\chi $,  the photon sphere radius is determined as
${{r}_{ps\alpha }}=6M\frac{1-\sqrt{1-\chi }}{\chi }$, In the context of a Schwarzschild black hole, with the same parameter, $\chi =\alpha \frac{48{{M}^{2}}}{{{L}^{2}}}$, the ratio of the photon sphere radius, ${{r}_{ps\alpha }}$, to the event horizon radius, ${{r}_{+}}$,
given by $\frac{{{r}_{ps\alpha }}}{{{r}_{+}}}=3\frac{1-\sqrt{1-\chi }}{\chi }$ is an invariant. Since $\chi =\alpha \frac{48{{M}^{2}}}{{{L}^{2}}}$, the EUP parameter $\alpha $ is inversely proportional to the square of the black hole mass $M$. That is, for larger black holes, the value of $\alpha $ is smaller, while for smaller black holes, the value of $\alpha $ is larger. This is in order to ensure that the ratio ${{r}_{ps\alpha }}/{{r}_{+}}$ remains constant. Conversely, for a fixed $\alpha $, we have $\chi \propto {{M}^{2}}$; consequently, as the black hole mass $M$ increase (decreases), the ratio ${{r}_{ps\alpha }}/{{r}_{+}}$ increase (decreases). In the limit $\chi \to 0$, the ratio attains its minimum value of $3\text{/}2$. In the limit $\chi \to 1$, the ratio ${{r}_{ps\alpha }}/{{r}_{+}}$ approaches its maximum value of $3$.


\begin{align}\label{3.11}
{{b}_{ps\alpha }}&=\frac{{\bar{L}}}{{{E}_{c}}}={{\left. \frac{r}{\sqrt{f(r,\alpha ,0)}} \right|}_{{{r}_{ps\alpha }}}}=\frac{{{r}_{ps\alpha }}}{\sqrt{\left( 1-\tfrac{2M}{{{r}_{ps\alpha }}} \right)\left( 1+\tfrac{4\alpha r_{ps\alpha }^{2}}{{{L}^{2}}} \right)}},
\end{align}
Substituting ${{r}_{ps\alpha }}=6M\frac{1-\sqrt{1-\chi }}{\chi }$ into Eq. (\ref{3.11}), we obtain
\begin{align}\label{3.12}
\frac{{{b}_{ps\alpha }}}{3M}&=\frac{2(1-\sqrt{1-\chi })}{\chi \sqrt{\left( 1-\frac{\chi }{3(1-\sqrt{1-\chi })} \right)\left( 1+\tfrac{3}{\chi }{{\left( (1-\sqrt{1-\chi }) \right)}^{2}} \right)}},
\end{align}

From Eq.(\ref{3.12}), it follows that as $\chi $ increase, ${{b}_{ps\alpha }}$ decreases. The maximum value is ${{b}_{ps\alpha }}=3\sqrt{3}M$, and the minimum value is ${{b}_{ps\alpha }}=\frac{3\sqrt{3}M}{2}$. When ${{b}_{ps\alpha }}={{r}_{ps\alpha }}$, the parameter $\chi $ satisfies
\begin{align}\label{3.13}
\sqrt{\left( 1-\frac{\chi }{3(1-\sqrt{1-\chi })} \right)\left( 1+\frac{3}{\chi }{{\left( (1-\sqrt{1-\chi }) \right)}^{2}} \right)}&=1,
\end{align}

Solving Eq. (\ref{3.13}) yields $\chi =0.75$. For $\chi =0.75$, we have ${{r}_{ps\alpha }}=6M\frac{1-\sqrt{0.25}}{0.75}=\frac{3M}{0.75}={{b}_{ps\alpha }}$. Since the shadow radius ${{b}_{ps\alpha }}$ and the photon sphere radius ${{r}_{ps\alpha }}$ must satisfy ${{b}_{ps\alpha }}\ge {{r}_{ps\alpha }}$, the allowed range is $0\le \chi \le 0.75$. Using Eqs. (\ref{3.10}) and (\ref{3.13}), we can plot the variations of the photon sphere radius ${{r}_{ps\alpha }}$ and the shadow radius ${{b}_{ps\alpha }}$ as functions of $\chi $.
\begin{figure}[htb]
\centering
\includegraphics[width=0.9\linewidth]{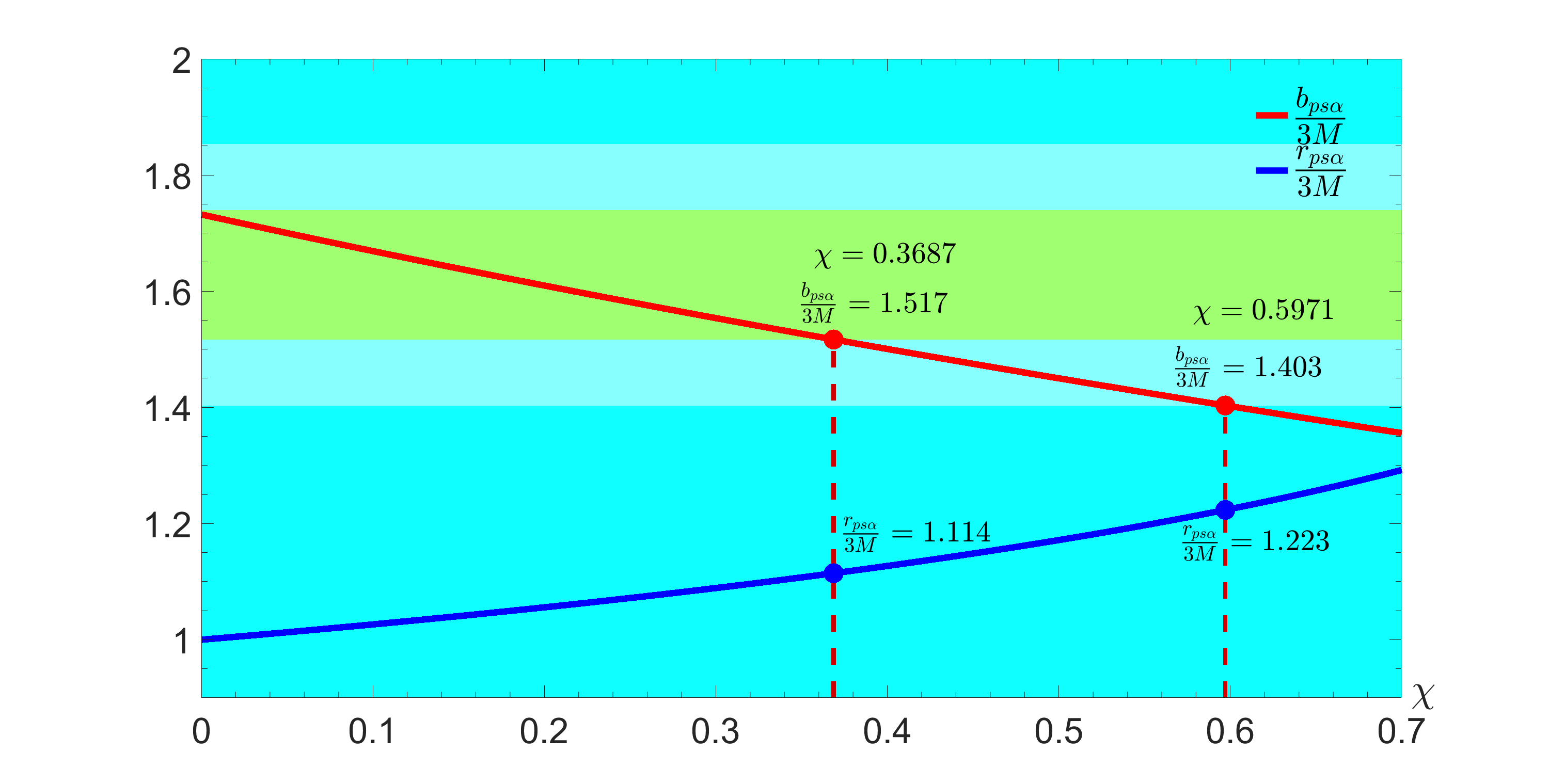}
\vskip -4mm
\caption{Curves of $\frac{b_{\it{{ps\alpha}}}}{3M}$ and $\frac{r_{\it{{ps\alpha}}}}{3M}$ as functions of $\chi$}
\label{fig3.1}
\end{figure}

Fig. \ref{fig3.1} shows the variations of the photon sphere radius ${{r}_{ps\alpha }}/3M$ and the critical impact parameter ${{b}_{ps\alpha }}/3M$ as functions of $\chi $. After averaging the Keck and VLTI mass-to-distance ratio priors for $\text{Sgr}{{\text{A}}^{*}}$, the yellow and white regions correspond to consistency with the EHT angular size images of $\text{Sgr}{{\text{A}}^{*}}$ at the 1$\sigma$ and 2$\sigma$ levels, respectively. Conversely, the cyan region is excluded by the same observations at more than 2$\sigma$.

As illustrated in Fig. \ref{fig3.1}, when the generalized uncertainty relation is taken into account, the photon sphere ${{r}_{ps\alpha }}/3M$ increases with $\chi $, while the shadow radius ${{b}_{ps\alpha }}/3M$ decreases as $\chi $ increases.
The constraint data from Keck and VLTI are as follows \cite{H250820419,S220507787,R220802969}:
\begin{align}\label{3.14}
4.55&\le {{b}_{ps\alpha }}\le 5.22,~~~~~~~~~within~1\sigma,\notag \\
4.21&\le {{b}_{ps\alpha }}\le 5.56,~~~~~~~~~within~2\sigma.
\end{align}

Under the constraints Eq. (\ref{3.14}). Fig. \ref{fig3.1} gives the upper bounds on $\chi $: ${{\chi }_{1}}=0.3687(1\sigma )$ and ${{\chi }_{2}}=0.597067(2\sigma )$. The corresponding photon sphere radii are ${{r}_{ps\alpha 1}}=3M\times 1.11449(1\sigma )$ and ${{r}_{ps\alpha 2}}=3M\times 1.22341(2\sigma )$, while the shadow radii ${{b}_{ps\alpha 1}}=3M\times 1.51687\approx 4.55M(1\sigma )$ and ${{b}_{ps\alpha 2}}=3M\times 1.40327\approx 4.21M(2\sigma )$. Thus, we obtain the allowed ranges for $\chi $: $0\le \chi \le 0.3687(1\sigma )$ and $0\le \chi \le 0.597067(2\sigma )$.

Using the relation $\chi =\alpha \frac{48{{M}^{2}}}{{{L}^{2}}}$, we can derive the allowed range for EUP parameter $\alpha /{{L}^{2}}$. From $\chi =\alpha \frac{48{{M}^{2}}}{{{L}^{2}}}$, it follows that, under the same observational constraints, the parameter $\alpha /{{L}^{2}}$ decreases as the black hole mass $M$ increases. In other words, the value of $\alpha /{{L}^{2}}$ deponds on the black hole mass $M$, and the EUP has a smaller impact on larger black holes.

Rewriting Eq. (\ref{3.4}) gives
\begin{align}\label{3.15}
{{V}_{eff}}(r,\alpha )&=\frac{\sqrt{\left( 1+\frac{\chi {{r}^{2}}}{12{{M}^{2}}} \right)\left( 1-\frac{2M}{r} \right)}}{r},
\end{align}
Setting $r=My$, we obtain
\begin{align}\label{3.16}
M{{V}_{eff}}(\chi ,\alpha )&=\frac{\sqrt{\left( 1+\frac{\chi }{12}{{y}^{2}} \right)\left( 1-\frac{2}{y} \right)}}{y},
\end{align}

\begin{figure}[htb]
\centering
\includegraphics[width=0.9\linewidth]{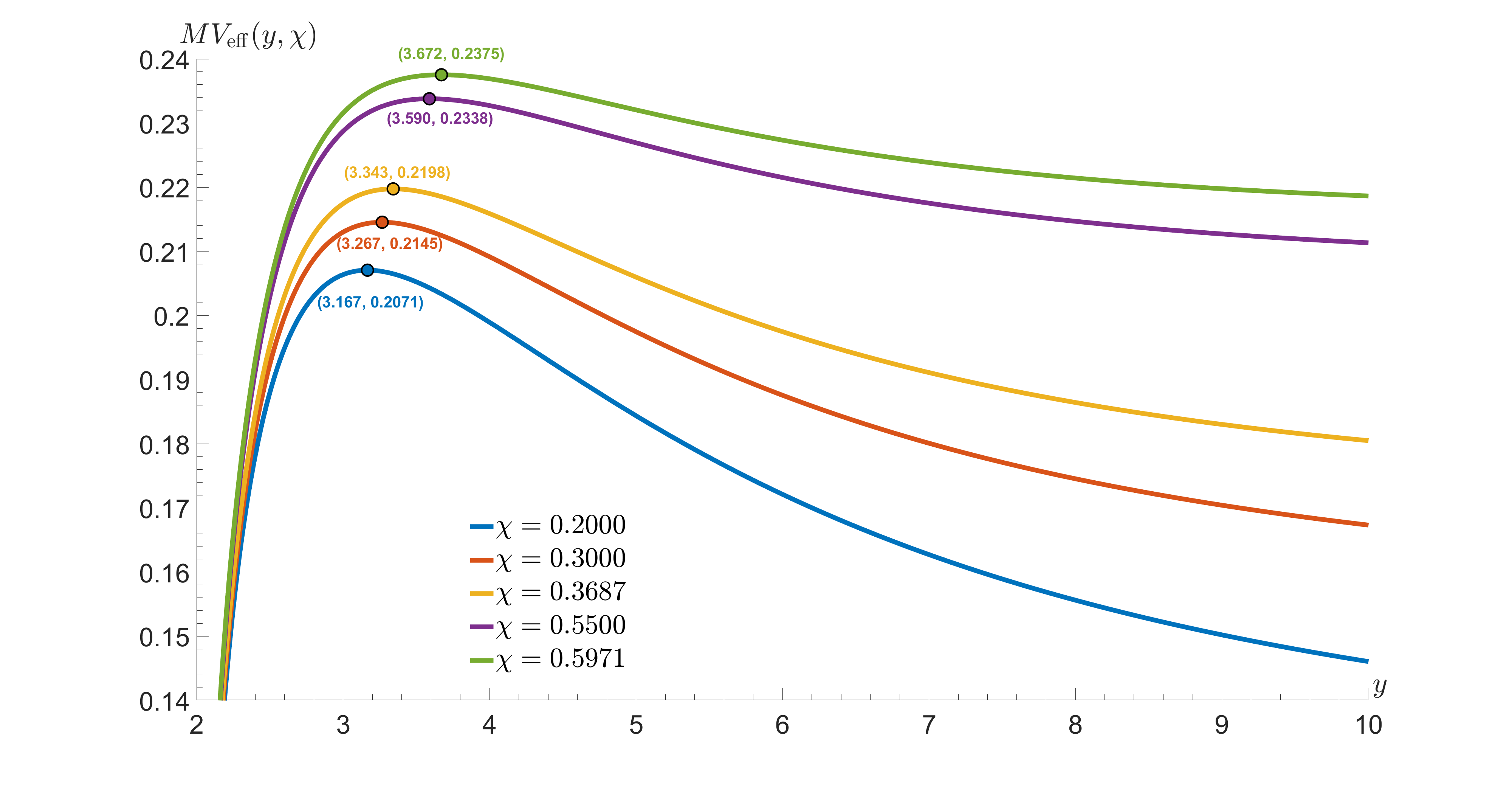}
\vskip -4mm
\caption{The $MV_{\text{eff}}-y$ curve}
\label{fig3.2}
\end{figure}

As shown in Fig. \ref{fig3.2}, with increasing $\chi $, the peak of the effective potential rises, and the photon sphere radius also increases. However, according to Fig. \ref{fig3.1}, the shadow radius decreases with increasing $\chi$. In contrast, for a typical black hole, the photon sphere radius and the shadow radius vary synchronously with changes in the effective potential ${{V}_{eff}}$: when ${{V}_{eff}}$ increases (or decreases), both the photon sphere radius and the shadow radius increase (or decrease) accordingly. Therefore, the influence of the EUP on the optical properties of black holes differs from other types of corrections \cite{H250820419}. The underlying reason for this distinct behavior warrants further investigation. This also opens up a new avenue for exploring how different modifications to the Hawking temperature affect the optical characteristics of black holes.

\section{discussion}\label{four}
It is well established that there exists a profound connection between gravity and thermodynamics, which may offer valuable insights into the microscopic origin of spacetime geometry. The identification of black holes as thermodynamic systems in possession of both temperature and entropy, has led to a growing realisation that gravitational dynamics can be conceptualised as an emergent phenomenon arising from underlying statistical degrees of freedom. Within this theoretical, the gravitational field equations can be interpreted as thermodynamic relations among quantities defined on the horizon. Consequently, modifications to the Hawking temperature are expected to induce corresponding modifications to the geometry itself.

Inspired by Refs. \cite{A251213769,S220507787,R220802969}, in this study, the object is to establish a self-consistent correspondence between the modified Hawking temperature and the spacetime metric when accounting for the Extended Uncertainty Principle (EUP) correction. This is based on the premise that the black hole satisfies the first law of thermodynamics. The objective of this study is to ascertain the spacetime metric that corresponds to the EUP-modified Hawking temperature. This approach provides a new avenue for further investigation of the thermodynamic and optical properties of black holes.

In Sec. \ref{two}, the line element of the spacetime incorporating EUP corrections is derived. The Einstein tensor was computed, yielding results analogous to those reported in Ref. \cite{A251213769}. It is evident that the effective matter sector originating from the EUP is anisotropic. In Sec. \ref{three}, the influence of the EUP on the photon sphere radius, the shadow radius, and the effective potential as functions of the introduced parameter, $\alpha /{{L}^{2}}$, is discussed. It has been established that the photon sphere radius increases with $\alpha /{{L}^{2}}$, whereas the shadow radius decreases as $\alpha /{{L}^{2}}$ increases.

Utilising the shadow measurements of $\text{Sgr}{{\text{A}}^{*}}$ from the Event Horizon Telescope (EHT), the upper bound of $\alpha /{{L}^{2}}$ has been constrained, thereby establishing a link between theoretical studies of the EUP and observational data. This process yields a permissible range for the uncertainty parameter $\alpha /{{L}^{2}}$ that has been introduced within the EUP framework.

The present study identifies several promising avenues for future research. One potential avenue for further research would be to extend the present analysis to rotating or charged black hole spacetimes under EUP-modified Hawking temperatures, and to explore their resulting optical properties, thereby enriching the scope of black hole physics. In addition, the anticipated enhancements in the precision of very long baseline interferometry (VLBI) observations will facilitate more precise constraints on the EUP parameter, $\alpha /{{L}^{2}}$, thereby leading to a refinement of the theoretical framework. This provides a timely opportunity to compare theoretical predictions with increasingly accurate astrophysical observations.

\section*{Acknowledgments}
We would like to thank Prof. Ren Zhao and Meng-Sen Ma for their indispensable discussions and comments. This work was supported by the Natural Science Foundation of China (Grant No. 12375050), Shanxi Provincial Natural Science Foundation of China (Grant No. 2025030221211241)

\end{document}